# SCMA for Downlink Multiple Access of 5G Wireless Networks

Hosein Nikopour, Eric Yi, Alireza Bayesteh, Kelvin Au, Mark Hawryluck, Hadi Baligh, and Jianglei Ma
Huawei Technologies Canada Inc., Ottawa, Ontario, Canada
{hosein.nikopour, zhihang.yi, alireza.bayesteh, kelvin.au, mark.hawryluck, hadi.baligh, jianglei.ma}@huawei.com

*Abstract*—Sparse code multiple access (SCMA) is a new frequency domain non-orthogonal multiple-access technique which can improve spectral efficiency of wireless radio access. With SCMA, different incoming data streams are directly mapped to codewords of different multi-dimensional cookbooks, where each codeword represents a spread transmission layer. Multiple SCMA layers share the same time-frequency resources of OFDMA. The sparsity of codewords makes the near-optimal detection feasible through iterative message passing algorithm (MPA). Such low complexity of multi-layer detection allows excessive codeword overloading in which the dimension of multiplexed layers exceeds the dimension of codewords. Optimization of overloading factor along with modulation-coding levels of layers provides a more flexible and efficient link-adaptation mechanism. On the other hand, the signal spreading feature of SCMA can improve link-adaptation as a result of less colored interference. In this paper a technique is developed to enable multi-user SCMA (MU-SCMA) for downlink wireless access. User pairing, power sharing, rate adjustment, and scheduling algorithms are designed to improve the downlink throughput of a heavily loaded network. The advantage of SCMA spreading for lightly loaded networks is also evaluated.

## I. INTRODUCTION

Future fifth generation (5G) wireless networks are expected to support better quality of service, higher throughput, and lower latency [1]. Recent research activities [2] [3] emerge toward new technologies providing solutions to meet the requirements of the next generation of wireless communication networks in 2020 and beyond.

Multi-user MIMO (MU-MIMO) is a well-known multiple access technique to share given time-frequency and power resources among multiple users in a downlink wireless access network [4]. The target is to increase the overall downlink throughput through user multiplexing. Multiple beams are formed over an array of antennas at a transmit point (TP) to serve multiple users distributed within a cell. Every MIMO layer is assigned to a user while layers are orthogonally separated in the space domain assuming MIMO beamforming precoders are properly selected according to the channels of target users. At the receive side, every user can simply match itself to its intended layer while other MIMO layers are seemed totally muted with no cross-layer interference, provided the precoders are properly designed. Despite the promising throughput gain and the simplicity of detection at user nodes, MU-MIMO as a closed-loop system suffers from some practical difficulties in terms of channel aging and high overhead to feed back channel state information (CSI) of users to a serving TP. CSI is required to form the best set of precoders for a selected set of paired users. If CSI is not well estimated, cross-layer interference practically limits the potential performance gain of MU-MIMO.

Open-loop user multiplexing is a desired approach to avoid practical limitations of MU-MIMO. Non-orthogonal code-domain multiple-access is an open-loop scheme to pair multiple users over shared time-frequency resources. Sparse code multiple access (SCMA) [5] [6] is a non-orthogonal codebook-based multiple-access technique with near optimal spectral efficiency. In SCMA, incoming bits are directly mapped to multi-dimensional complex codewords selected from predefined codebook sets. Co-transmitted spread data are carried over super-imposed layers.

SCMA is well-matched to user multiplexing as we can allocate code-domain layers to different users without need for CSI knowledge of paired users. In this paper, multi-user SCMA (MU-SCMA) is proposed to improve a network throughput. With a very limited need for channel knowledge in terms of channel quality indicator (CQI), TP simply pairs users together while the transmit downlink power is properly shared among multiplexed layers. Compared to MU-MIMO, this system is more robust against channel variations. In addition, the problem of CSI feedback is totally removed for this open-loop multiple-access scheme.

Since layers are not fully separated in a non-orthogonal multiple access system, a non-linear receiver is required to detect the intended layer of every user. Therefore, further complexity of detection is the cost of the non-orthogonal multiple-access especially when a system is heavily overloaded with a large number of multiplexed layers. Sparsity of SCMA codewords lets us take advantage of the low complexity message passing algorithm (MPA) [7] detector with ML-like performance. MPA performs well even if the system is overloaded with a large number of layers.

Low density spreading (LDS) [7] is a special form of SCMA. In LDS, codewords are built by spreading of modulated QAM symbols using low-density spreading signatures with a few numbers of nonzero elements within a large signature length. Despite the moderate complexity of detection, LDS suffers from poor performance especially for large constellation sizes above QPSK. All CDMA schemes and in particular LDS can be considered as different types of repetition coding in which different variations of a QAM symbol are generated by a spreading signature. Repetition coding is not able to provide desirable spectral efficiency for a wide range of SNR. To overcome this problem, in SCMA the QAM mapper and linear operation of sparse spreading are merged together to directly map incoming bits to a complex sparse vector called a codeword. This enables SCMA to benefit from shaping gain [8] of multi-dimensional constellations as opposed to simple repetition coding of linear sparse sequences. Consequently, SCMA substantially improves spectral efficiency of linear sparse sequences through multi-dimensional shaping gain of codebooks while it still provides other benefits in terms of overloading and moderate complexity of detection.

Interference management and robustness of the link quality are concerns in lightly loaded networks. When the traffic demand is low, the resource utilization drops. Within the long-term evolution (LTE) [9] context, as an example of an OFDMA system, that is equivalent to muting of some resource blocks (RB) across the bandwidth of a TP. In this situation, the interference level at a downlink user changes rapidly within every scheduling interval even if fading channels are quite stable with very slow variation. Interference level at an RB rises if most of RBs of neighboring cells are occupied, and drops if corresponding RBs of neighboring cells are empty. This rapid variation of interference level in time and frequency is not predictable in practice when there is no dynamic cooperation among the neighboring cells. The system has no choice but to adapt itself to the worst case scenario of channel quality. Poor link-adaptation inherently decreases the efficiency of a link.

Spreading over OFDMA tones can potentially improve the quality of link-adaptation procedure due to the interference averaging [10]. By using SCMA spreading technique, interference from different TPs is averaged out over the spread tones. This makes the interference white which has the advantage of better and more robust link-adaptation. In addition, layer multiplexing adds another degree of freedom to the link-adaptation capability of an SCMA system. Number of layers along with codebook sizes, coding rates, and power level of multiplexed layers are the parameters dictating the rate and quality of a link.

This paper evaluates the advantage of SCMA in a downlink wireless network. Two scenarios are considered: i) a fully loaded network with high throughput demand, and ii) a lightly loaded network with a fast variation of interference within each scheduling interval. MU-SCMA is proposed and evaluated in a heavily loaded network for the sake of throughput improvement. The techniques related to MU-SCMA are developed to pair users over shared time-frequency resources. The impact of interference averaging due to SCMA spreading is also evaluated for a lightly loaded network scenario.

Throughout this paper, $\mathbf{x}$ is a vertical vector, $\mathbf{X}$ represents a matrix, $\mathbf{1}_N$ is an all-one vector of size $N$, and $\mathbf{I}_N$ depicts an $N \times N$ identity matrix.

The rest of the paper is organized as follows. Section II defines the SCMA system model and structure. Section III is dedicated to the algorithms required to enable MU-SCMA. This section describes the details of scheduling and user paring algorithms, rate adjustment, detection strategies and power allocation of the paired users. Numerical results are reported in Section IV. The paper finally concludes in Section V.

## II. SYSTEM MODEL AND DESCRIPTION

### A. Downlink SCMA Model

An SCMA encoder is defined as a mapping from $\log_2(M)$ coded bits to a $K$-dimensional complex codebook of size $M$. The $K$-dimensional complex codewords of the codebook are sparse vectors with $N < K$ non-zero entries. All the codewords in the codebook contain 0 in the same $K - N$ dimensions.

An SCMA encoder contains $U$ users each with $J_u$, $u = 1, \ldots, U$ separate layers. SCMA codewords are multiplexed over $K$ shared orthogonal resources, e.g. OFDMA tones. In a downlink single-input multiple-output (SIMO) channel, the received signal of antenna $r$ of user $u_0$ can be expressed as

$$\mathbf{y}_{ru_0} = \mathrm{diag}(\mathbf{h}_{ru_0}) \sum_{u=1}^{U} \sqrt{p_u/J_u} \sum_{j=1}^{J_u} \mathbf{x}_{ju} + \mathbf{n}_{ru_0}, \quad (1)$$

where $J_u$ is the number of layers allocated to user $u$, $\mathbf{x}_{ju}$ is the $j$th SCMA codeword of user $u$ such that $\|\mathbf{x}_{ju}\|^2 = K$, and $p_u$ is the total transmit power per tone allocated to user $u$. The power of user $u$ is equally distributed among $J_u$ layers. The total transmit power is $P = \sum_{u=1}^{U} p_u$ and the total number of multiplexed layers is $J = \sum_{u=1}^{U} J_u$. The channel vector of $r$th receive antenna of user $u$ is $\mathbf{h}_{ru}$ in which every element represents the channel of an OFDMA tone. The ambient noise vector of user $u$ at receive antenna $r$ is represented by $\mathbf{n}_{ru}$. For the sake of notation simplification, adjacent OFDMA tones can be assumed identical, i.e. $\mathbf{h}_{ru} = h_{ru}\mathbf{1}_K$.

As SCMA is a non-linear modulator, it is not straightforward to model and extract its capacity. Hence, in the rest of the paper the linear sparse sequence modeling is used to develop the related pairing algorithms for MU-SCMA. In the sequel, the MIMO equivalent model of a linear sparse sequence system is developed.

### B. MIMO Equivalent of Linear Sparse Sequence

A linear sparse sequence is simply the spread version of a QAM symbol such that $\mathbf{x}_{ju} = \mathbf{s}_{ju} q_{ju}$ where $\mathbf{s}_{ju}$ is the $j$th signature vector of user $u$ such that $\|\mathbf{s}_{ju}\|^2 = K$ and $q_{ju}$ is its corresponding QAM symbol with unit average power. Let $\mathbf{S}_u = (\mathbf{s}_{1u}, \ldots, \mathbf{s}_{J_u u})$ represent the signature matrix of user $u$, and $\mathbf{q}_u = (q_{1u}, \ldots, q_{J_u u})^T$. (1) can be rewritten as

$$\mathbf{y}_{ru_0} = h_{ru_0} \sum_{u=1}^{U} \sqrt{p_u/J_u} \mathbf{S}_u \mathbf{q}_u + \mathbf{n}_{ru_0}. \quad (2)$$

Stacking received signal of all $R$ antennas together, the linear sparse sequence model is expressed as a MIMO system

$$\mathbf{y}_{u_0} = \sum_{u=1}^{U} \sqrt{p_u/J_u} \mathbf{H}_{u_0 u} \mathbf{q}_u + \mathbf{n}_{u_0}, \quad (3)$$

where $\mathbf{y}_u = (\mathbf{y}_{1u}^T, \ldots, \mathbf{y}_{Ru}^T)^T$, $\mathbf{n}_u = (\mathbf{n}_{1u}^T, \ldots, \mathbf{n}_{Ru}^T)^T$ and $\mathbf{H}_{u_0 u} = \mathbf{h}_{u_0} \otimes \mathbf{S}_u$ in which $\mathbf{h}_u = (h_{1u}, \ldots, h_{Ru})^T$ and $\otimes$ represents Kronecker product. By multiplexing $J$ layers over $K$ resources, the overloading factor of the system is $J/K$.

Assuming all $J$ layers belong to one user, i.e. $U = 1$, the open-loop capacity of a MIMO system can be expressed as

$$C = \log_2 \det(\mathbf{I}_{RK} + P/J\mathbf{H}^H\mathbf{R}_{nn}^{-1}\mathbf{H}), \quad (4)$$

where $\mathbf{R}_{nn}$ is the covariance matrix of noise $\mathbf{n}$. Note that index $u$ is dropped for the simplicity of the notation. This is the upper bound rate of a single-user sparse spreading sequence system with a given channel, signature matrix, and power allocation of layers. This bound is tight for QPSK modulation but deviates from performance of linear sparse sequence for higher constellation sizes where the repetition coding fails. The benefit of SCMA is to recover this performance gap through multi-dimensional modulation.

Assuming $\mathbf{R}_{nn} = N_1 \mathbf{I}_{RK}$, (4) is equivalent to

$$C = \log_2 \det(\mathbf{I}_{RK} + P/JN_1\mathbf{H}^H\mathbf{H}). \quad (5)$$

Based on the definition of $\mathbf{H}$, one can simply show that

$$\mathbf{H}^H\mathbf{H} = \|\mathbf{h}\|^2(\mathbf{S}^H\mathbf{P}\mathbf{S}). \quad (6)$$

By replacing (6) into (5), we have

$$C = \log_2 \det(\mathbf{I}_{RK} + \gamma/J(\mathbf{S}^H\mathbf{S})), \quad (7)$$

in which $\gamma := \|\mathbf{h}\|^2 P/N_1$ is the instantaneous SIMO SNR of the user.

According to (7), the rate of a single-user system depends on SNR as well as the number of layers and the signature matrix. Consequently, it can provide further degree of freedom for link-adaptation of an SCMA system. In the case of OFDMA with $\mathbf{S} = (1)$, $J = 1$, and $U = 1$, (7) is simply reduced to OFDMA SIMO capacity, i.e. $C = \log_2(1 + \gamma)$.

## III. DOWNLINK MU-SCMA ALGORITHMS

In downlink MU-SCMA, the transmitted layers belong to more than one user. Several algorithms are required to enable MU-SCMA. Paired users are first selected from a pool of users. The user selection criterion and its optimization approach are part of the design parameters. Signals of paired users are transmitted from an antenna with a total power constraint. Therefore, the power has to be split among paired users according to their channel conditions. Following the power allocation, the rate of each user is adjusted to match the target error rate and link quality. Codebook size, coding rate and number of layers are the parameters to adjust the rate of each paired user.

### A. User Pairing to Maximize Wiegthed Sum-Rate

Assume a pool of users with instantaneous SINR $\gamma_u = \|\mathbf{h}_u\|^2 P/N_u$, $u = 1, \ldots, U$, and average rates $R_u$. In the absence of user pairing, a proportional fair (PF) scheduler maximizes the weighted rate as follows

$$u^* = \max_u \frac{r_u}{R_u}, \quad (8)$$

where $r_u$ is determined according to the available CQI, e.g. $r_u = \log_2(1 + \gamma_u)$ for OFDMA.

The target of pairing is to maximize the weighted sum-rate. Assuming two users, the weighted sum-rate is expressed as

$$u_1^*, u_2^* = \max_{u_1, u_2} \frac{\tilde{r}_{u_1}}{R_{u_1}} + \frac{\tilde{r}_{u_2}}{R_{u_2}}, \quad (9)$$

in which $\tilde{r}_{u_1} := \tilde{r}_{u_1}(u_1, u_2; \text{power sharing})$ is the adjusted rate of user $u_1$ after pairing. Notably, the adjusted rate depends on both paired users and the power sharing strategy. Following the exhaustive search, all $U(U - 1)$ pairing options as well as $U$ single-user options must be checked to find the solution of (9). The complexity of the exhaustive search increases in the order of $U^2$, which might not be practically feasible.

A greedy algorithm can be used to reduce the complexity of pairing. In a greedy approach, the first user is picked according to the single-user scheduling criterion of (8) and then another user is paired with the first one. With greedy scheduling, the complexity order of two user pairing is reduced to $U - 1$. The steps of the greedy scheduling is listed as below

$$u_1^* = \max_u \frac{r_u}{R_u}, \quad (10)$$

$$u_2^* = \max_{u \neq u_1^*} \frac{\tilde{r}_{u_1^*}}{R_{u_1^*}} + \frac{\tilde{r}_u}{R_u}, \quad (11)$$

and the pairing result is valid only if the following condition is satisfied

$$\frac{\tilde{r}_{u_1^*}}{R_{u_1^*}} + \frac{\tilde{r}_{u_2^*}}{R_{u_2^*}} > \frac{r_{u_1^*}}{R_{u_1^*}}. \quad (12)$$

### B. Rate Adjustment and Detection Strategy

Assuming $U = 2$, $p_1 = \alpha P$, and $p_2 = (1 - \alpha)P$, (3) can be rewritten as follows for two users

$$\mathbf{y}_u = \sqrt{\frac{\alpha P}{J_1}}\mathbf{H}_{u1}\mathbf{q}_1 + \sqrt{\frac{(1-\alpha)P}{J_2}}\mathbf{H}_{u2}\mathbf{q}_2 + \mathbf{n}_u, u = 1,2, \quad (13)$$

in which $\alpha \in (0, 1.0)$ represents the power sharing factor. Without loss of generality, we assume $\gamma_1 > \gamma_2$.

The above equation models a MIMO broadcast channel which is not generally degraded [11]. In other words, the assumption of $\gamma_1 > \gamma_2$ does not necessarily imply that user 1 can decode user 2's data and achieve a higher rate compared to this user. For the sake of simplifying the analysis, the model is transformed to an approximated degraded model. The signal of user 1 is treated as interference at user 2 such that the equivalent interference is described as

$$\mathbf{R}_2 = N_2\mathbf{I}_{RK} + \alpha P/J_1\mathbf{H}_{21}\mathbf{H}_{21}^H. \quad (14)$$

It can be shown that $\mathbf{H}_{21}\mathbf{H}_{21}^H = (\mathbf{h}_2\mathbf{h}_2^H) \otimes (\mathbf{S}_1\mathbf{S}_1^H)$ meaning that equivalent interference at user 2 is colored even if the background noise in white. According to (7), the broadcast model of (13) becomes degraded if the interference in (14) is approximated with a white noise as follows

$$\mathbf{R}_2 \approx (N_2 + \alpha P\|\mathbf{h}_2\|^2)\mathbf{I}. \quad (15)$$

In other words, with the above approximation, if user 2 is able to decode its own data, user 1 can also decode it.

For the sake of simplicity, the intended algorithms are first developed for OFDMA (which is degraded by nature) and then extended to SCMA. For OFDMA, (13) is rewritten as follows

$$\mathbf{y}_u = \mathbf{h}_u\sqrt{\alpha P}q_1 + \mathbf{h}_u\sqrt{(1 - \alpha)P}q_2 + \mathbf{n}_u. \quad (16)$$

As $\gamma_1 > \gamma_2$, in the degraded OFDMA channel, user 1 is the good quality user with a higher rate compared to user 2 with a lower instantaneous rate.

*1) Detection at high-quality user 1*

The capacity region of user 1 is shown in Fig. 1 with solid lines. An ideal joint ML detector is feasible at user 1 if

$$\tilde{r}_1 + \tilde{r}_2 \leq \log_2(1 + \gamma_1), \quad (17)$$

subject to

$$\tilde{r}_2 \leq \log_2(1 + \tilde{\gamma}_{2@1}), \quad (18)$$

and

$$\tilde{r}_1 \leq \log_2(1 + \tilde{\gamma}_1). \quad (19)$$

Referring to (16) for $u = 1$, $\tilde{\gamma}_{2@1}$ represents the effective SINR for the single-user detection of user 2 at user 1 while user 1 is treated as an interferer, i.e.

$$\tilde{\gamma}_{2@1} = \frac{\|\mathbf{h}_1\|^2(1-\alpha)P}{\|\mathbf{h}_1\|^2\alpha P + N_1} = \frac{(1-\alpha)\gamma_1}{1+\alpha\gamma_1}. \quad (20)$$

Condition (18) guarantees the single-user detection of the low rate user 2 at the good quality user 1. Assuming user 2 is totally decodable at user 1, after successive interference cancellation (SIC), the detection problem of user 1 is reduced to $\mathbf{y}_1 = \mathbf{h}_1\sqrt{\alpha P}q_1 + \mathbf{n}_1$. In this case, user 1 is decodable if condition (19) is guaranteed in which

$$\tilde{\gamma}_1 = \frac{\|\mathbf{h}_1\|^2\alpha P}{N_1} = \alpha\gamma_1. \quad (21)$$

The SIC detection strategy corresponds to point C in the capacity region of Fig. 1 where the maximum achievable rate is provided for user 1.

*2) Detection at low-quality user 2*

Since user 2 has lower channel quality, it is not able to detect user 1 with a higher rate. Therefore, user 2 has no choice but to treat user 1 as its co-paired interference. In the other hand, the capability of user 1 to detect user 2 does not necessarily mean that user 2 can also detect its own intended data. Consequently, a tighter condition on the rate of user 2 might be required to make the detection of user 2 signal possible for its target user. Referring to (16) for $u = 2$, the SINR for single-user detection of user 2 is expressed as

$$\tilde{\gamma}_2 = \frac{\|\mathbf{h}_2\|^2(1-\alpha)P}{\|\mathbf{h}_2\|^2\alpha P + N_2} = \frac{(1-\alpha)\gamma_2}{1+\alpha\gamma_2}. \quad (22)$$

Therefore, the signal of user 2 is detectable at user 2 with single-user detection only if

$$\tilde{r}_2 \leq \log_2(1 + \tilde{\gamma}_2). \quad (23)$$

*3) Optimum operating point*

The capacity region of user 1 and 2 are illustrated jointly in Fig. 1. The shaded region corresponds to an area in which every user can detect its intended data stream while the power sharing factor is $\alpha$. The shaded area satisfies conditions (17) to (19) as well as (23).

Theoretically, the best point of transmission is either A or B of Fig. 1 depending on the weights of the users. The weighted sum-rate at point B is described as

$$WSR_B(\alpha) = \frac{\log_2\left(1+\frac{\alpha\gamma_2}{1+(1-\alpha)\gamma_2}\right)}{R_1} + \frac{\log_2(1+(1-\alpha)\gamma_2)}{R_2}. \quad (24)$$

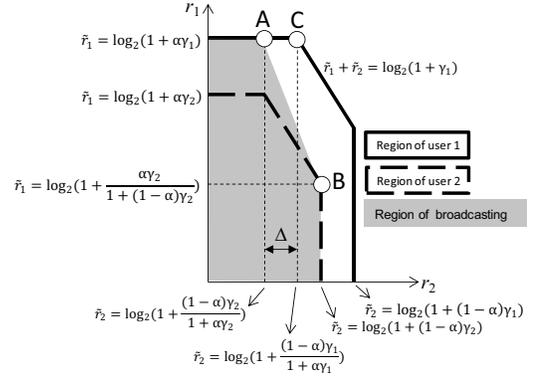

Fig. 1. Capacity regions at user 1 and 2.

Assuming $R_1 > R_2$, the weighted sum-rate at B is maximized only if $\alpha = 0$. Under this condition, the scheduling result falls back to the single-user scheduling. Therefore, if the user pairing provides better weighted sum-rate, the only promising scenario of interest is point A.

Assuming the pairing providers gain at point A, the quality of single-user detection of user 2 signal at user 1 is much better than detection of user 2 signal at user 2. It can be shown by comparing $\tilde{\gamma}_{2@1}$ and $\tilde{\gamma}_2$. Referring to (20), (22), and Fig. 1, the detection margin of user 2 is defined as

$$\Delta := \frac{\tilde{\gamma}_{2@1}}{\tilde{\gamma}_2} = \frac{\gamma_1}{\gamma_2}\frac{1+\alpha\gamma_2}{1+\alpha\gamma_1}. \quad (25)$$

*C. OFDMA Power Sharing Optimization*

The goal is to optimize the power sharing factor $\alpha$ while the two paired users operate at point A of Fig. 1. The weighted sum-rate of the paired users at point A is

$$WSR_A(\alpha) = \frac{\log_2(1+\alpha\gamma_1)}{R_1} + \frac{\log_2\left(1+\frac{(1-\alpha)\gamma_2}{1+\alpha\gamma_2}\right)}{R_2} \quad (26)$$

The optimum $\alpha$ is the solution of $\frac{\partial WSR_A(\alpha)}{\partial \alpha} = 0$ and can be easily derived as

$$\alpha^* = \frac{R_1\gamma_2 - R_2\gamma_1}{(R_2-R_1)\gamma_1\gamma_2}. \quad (27)$$

The optimal solution $\alpha^*$ is valid only if $\alpha^* \in (0,1)$. To facilitate the detection, one may limit $\alpha$ to a smaller range.

Referring to (25) and (27), the detection margin of users 2 for optimal $\alpha = \alpha^*$ is reduced to

$$\Delta = \frac{\gamma_1}{\gamma_2}\frac{1+\alpha^*\gamma_2}{1+\alpha^*\gamma_1} = \frac{R_1}{R_2}. \quad (28)$$

It explains that two paired users are easily separated if the ratio of their long-term rates is large enough.

*D. SCMA Power Sharing Optimization*

Following the same detection strategy as OFDMA, by referring to (7) and (21), the adjusted rate of the first paired user can be described as

$$\tilde{r}_1 = \log_2 \det\left(\mathbf{I} + \frac{\alpha\gamma_1}{J_1}(\mathbf{S}_1^H\mathbf{S}_1)\right). \quad (29)$$

Referring to (7), (13), and (15), the adjusted rate of second user can be expressed as

$$\tilde{r}_2 = \log_2 \det(I + \frac{(1-\alpha)P\|h_2\|^2}{J_2(N_2+\alpha P\|h_2\|^2)}(S_2^H S_2)), \quad (30)$$

or equivalently

$$\tilde{r}_2 = \log_2 \det(I + \frac{(1-\alpha)\gamma_2}{J_2(1+\alpha\gamma_2)}(S_2^H S_2)). \quad (31)$$

The weighted sum-rate with weights $w_i = 1/R_i$ is

$$WSR_A(\alpha) = w_1 \log_2 \det(I + \frac{\alpha\gamma_1}{J_1}(S_1^H S_1)) + \quad (32)$$
$$w_2 \log_2 \det(I + \frac{(1-\alpha)\gamma_2}{J_2(1+\alpha\gamma_2)}(S_2^H S_2)).$$

The Hermitive matrix $S_u^H S_u$ can be decomposed to $U_u \Lambda_u U_u^H$, and hence $\log_2 \det(I + \rho S_u^H S_u) = \sum_i \log_2(1 + \rho\lambda_{ui})$ in which $\lambda_{ui}$ is the $i$th diagonal element $\Lambda_u$. Rewriting (32), we have

$$WSR_A(\alpha) = w_1 \sum_{i=1}^{\min(K,J_1)} \log_2(1 + \frac{\alpha\gamma_1}{J_1}\lambda_{1i}) + \quad (33)$$
$$w_2 \sum_{i=1}^{\min(K,J_2)} \log_2(1 + \frac{(1-\alpha)\gamma_2}{J_2(1+\alpha\gamma_2)}\lambda_{2i}).$$

The optimum power sharing factor is the solution of $\frac{\partial WSR_A(\alpha)}{\partial \alpha} = 0$ which leads to the following

$$\sum_{i=1}^{\min(K,J_1)} \frac{w_1 \gamma_1 \lambda_{1i}}{J_1 + \gamma_1 \lambda_{1i}\alpha^*} - \quad (34)$$
$$\sum_{i=1}^{\min(K,J_2)} \frac{w_2 \gamma_2 \lambda_{2i}(1+\gamma_2)}{(1+\gamma_2\alpha^*)(J_2+\gamma_2\lambda_{2i}+(J_2-\lambda_{2i})\gamma_2\alpha^*)} = 0.$$

The solution of the above polynomial is $\alpha^*$ which is valid only if it is real and belongs to the interval (0,1.0). In the case of multiple solutions, the one that maximizes $WSR_A(\alpha^*)$ is selected.

## IV. NUMERICAL RESULTS

Simulations are done to evaluate the benefits of SCMA and MU-SCMA in a downlink wireless cellular network. As listed in TABLE I, the common simulation assumptions and parameters follow the 3rd generation partnership project (3GPP) evaluation methodology [12]. The codebooks of SCMA are designed based on the principles reported in [6]. The dimension of SCMA codewords is 4 with 2 non-zero elements in each codeword. The maximum number of SCMA layers is 6. SCMA layers are detected with the near-optimal MPA detector and the detection strategy for MU-SCMA follows the approach described in Section III.

The system-level simulation results are listed in TABLE II comparing OFDMA with SCMA and MU-SCMA in terms of cell aggregate throughput and 5 percentile coverage rate of cell edge users. The traffic model is full buffer with average 10 users per cell in 10 MHz LTE network. This scenario is equivalent to a heavily loaded network with a high demand for throughput. The scheduler is PF with wideband scheduling meaning that a scheduled user occupies the whole resources within a scheduling interval.

Extracting from the results of TABLE II, the relative throughput and coverage gains of SCMA and MU-SCMA with respect to OFDMA are illustrated in 0.

TABLE I. SIMULATION ASSUMPTIONS AND PARAMETERS FOR NETWORK SCENARIOS

| Parameter | Value |
|---|---|
| Deployment | Hexagonal grid, 19 sites, 3 sectors per site, and 500 m inter-site distance |
| Distance-dependent path loss | $L = 128.1 + 37.6 \log_{10}(D)$, $D$ in km and $D > 35$ m |
| Penetration loss | 20 dB |
| Shadowing | 8 dB long-normal shadowing |
| TP antenna pattern | 3GPP 3D model |
| Number of users | 570 users uniformly distributed across the entire network |
| System bandwidth | 10 MHz at 2 GHz carrier frequency |
| Channel type | 1×2 ITU-TU fading channel |
| User speed | 3 and 30 km/h |
| HARQ | Incremental redundancy (IR) HARQ with up to 3 retransmissions |
| Scheduler | PF with wideband or subband scheduling |
| SCMA codeword dimension | 4 OFDMA tones |
| SCMA codebook sizes | 4, 8, and 16 |
| SCMA maximum number of layers | 6 with overloading factor of 1.5 |
| SCMA receiver | MPA joint detector |

TABLE II. SYSTEM SIMULATION RESULTS COMPARING OFDMA, SCMA, AND MU-SCMA FOR A FULL BUFFER SCENARIO WITH WIDEBAND PF SCHEDULING

| Radio Access Mode | Throughput [Mbps] | Coverage [Kbps] |
|---|---|---|
| OFDMA | 18.6 | 574.7 |
| SCMA | 19.6 | 621.8 |
| MU-SCMA | 24.0 | 779.7 |

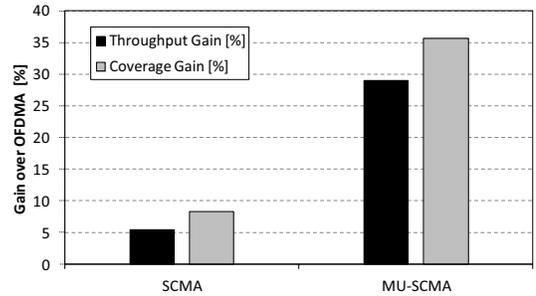

Cell throughput and coverage gain of SCMA and MU-SCMA over OFDMA. According to 0, SCMA shows 5% throughput and 8% coverage gain over OFDMA. The major source of the gain is multi-dimensional shaping gain of SCMA codebooks and the flexibility of the link-adaptation associated with SCMA. However, the major advantage of SCMA over OFDMA appears when the non-orthogonal MU-SCMA technique is implemented. Referring to 0, the throughput and coverage gains of MU-SCMA over OFDMA are 28% and 36%, respectively. The reported gain here is for users with 3 km/h speed, but the same results are also derived for a network with higher user speeds. Therefore, the robustness of open-loop MU-SCMA to channel variation is one of its main advantages over MU-MIMO.

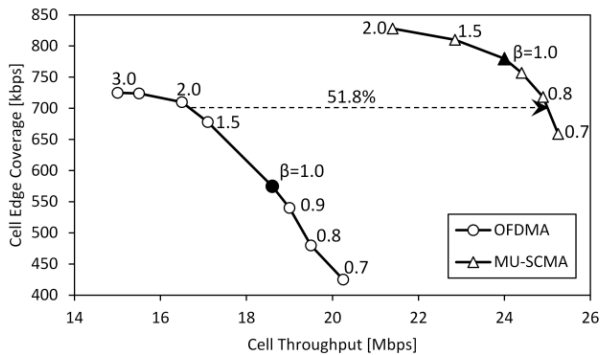

Fig. 2. Cell coverage and throughpurt trade-off for OFDMA and MU-SCMA. For coverage rate at 700 kbps, the gain of cell throughput is more than 51%.

The metric of a PF scheduler can be modified to establish a trade-off between coverage and throughput of a network. The weight of the PF metric is modified to $w_u = 1/R_u^\beta$ where $\beta$ is the exponent of the average rate to adjust the outcome of the original weighted sum-rate of (32). Fig. 2 depicts how $\beta$ impacts the coverage and throughput rates of OFDMA and MU-SCMA transmission modes. For a given coverage rate at 700 kbps, the throughput gain of MU-SCMA with respect to OFDMA is more than 51%.

The results of the evaluation of a network with 50% resource utilization are listed in TABLE III to show the advantage of SCMA over OFDMA in terms link-adaptation in a highly variant interference environment. In an OFDMA system, in average 50% of RBs are randomly off within a scheduling interval. Lower resource utilization reduces the average interference level but increases the time and frequency variation of the interference power.

In SCMA the whole bandwidth might be allocated but 50% of layers are not randomly assigned during the subband PF scheduling. The width of each subband is 5 RBs. Despite OFDMA in which an RB is either on with full power or completely off, the transmit power of an RB in SCMA changes according to its corresponding number of active layers. It alleviates the rapid variation of interference across RBs in the frequency direction.

TABLE III. SYSTEM SIMULATION RESULTS COMPARING OFDMA AND SCMA WITH 50% RESOURCE UTILIZATION AND SUBBAND PF SCHEDULING

| Radio Access Mode | Throughput [Mbps] | Coverage [Kbps] |
|---|---|---|
| OFDMA | 13.0 | 526.3 |
| SCMA | 20.3 (56.2%) | 665.7 (26.5%) |

According to TABLE III, the throughput and coverage gains of SCMA over OFDMA are 56% and 26%, respectively. Therefore, SCMA can bring robustness to a lightly-loaded system through interference averaging due to slow variation of transmit power across frequency and the spreading of codewords across multiple OFDMA tones.

## V. CONCLUSION

MU-SCMA is introduced to increase the downlink spectral efficiency of 5G wireless cellular networks. SCMA is a non-orthogonal multiple-access scheme. User multiplexing is realized without need for a full knowledge of users' instantaneous channels. This feature provides an advantage for MU-SCMA over other existing multiplexing techniques such as MU-MIMO in which sensitivity to channel aging and high overhead of channel knowledge feedback are the obstacles for their practical implementation in a real network. In addition, compared to MU-MIMO schemes which are based on spatial domain precoding, code-domain multiplexing has a substantial advantage in terms of the transmit side computational complexity. Promising performance gain of MU-SCMA makes it attractive for future wireless networks. In this paper, MU-SCMA techniques are developed for a single-TP and SIMO channel. The extension of MU-SCMA to multiple-TP and MIMO scenarios is a direction for future research activities.